\documentclass{osa-article-arxiv}

\journal{oe}
\usepackage{todonotes}
\usepackage{soul}
\usepackage{color}

\newcommand{\bd}{\begin{displaymath}}
\newcommand{\ed}{\end{displaymath}}
\newcommand{\be}{\begin{equation}}
\newcommand{\ee}{\end{equation}}
\newcommand{\ba}{\begin{eqnarray}}
\newcommand{\ea}{\end{eqnarray}}

\articletype{Research Article}
\begin{document}

\title{Uncertainty principle for axial power content of highly focused fields}

\author{R. Mart{\'\i}nez-Herrero,\authormark{1} A. Carnicer,\authormark{2} I. Juvells,\authormark{2} and A. S. Sanz\authormark{1,*}}

\address{\authormark{1}Department of Optics, Faculty of Physical Sciences, Universidad Complutense de Madrid, Pza.\ Ciencias 1, Ciudad Universitaria -- 28040 Madrid, Spain\\
\authormark{2}Universitat de Barcelona (UB), Facultat de F{\'\i}sica, Departament de F{\'\i}sica Aplicada, Mart{\'\i} i Franqu{\`e}s 1, 08028 Barcelona, Spain}

\email{\authormark{*}a.s.sanz@fis.ucm.es} %% email address is required



\begin{abstract}
In the analysis of the on-axis intensity for a highly focused optical field it is highly desirable to deal with effective relations aimed at characterizing the field behavior in a rather simple fashion. Here, a novel and adequate measure for the size of the region where the axial power content mainly concentrates is proposed on the basis of an uncertainty principle. Accordingly, a meaningful relationship is provided for both the spread of the incident beam at the entrance of the highly focused optical system and the size of the region where the on-axis power mainly concentrates.
\end{abstract}

%%%%%%%%%%%%%%%%%%%%%%%%%%  body  %%%%%%%%%%%%%%%%%%%%%%%%%%

\section{Introduction}
\label{sec1}

Based on the principles of classical optics, Heisenberg explained one of the most fundamental results of quantum mechanics, namely the uncertainty principle, in term of what is nowadays known as the Heisenberg microscope \cite{heisengerg:ZPhys:1927,zurek-bk}. Ever since, different generalizations to the uncertainty relations have been considered in the literature \cite{chisolm:AJP:2001,busch:PhysRep:2007,maccone:PRL:2014}, with extensions even in fields such as string theory \cite{veneziano:EPL:1986}, quantum gravity \cite{kempf:PRD:1995,scardigli:PhysLettB:1999}, or corpuscular gravity \cite{buoninfante:EurPhysJC:2019}. Notwithstanding its applicability has also be subjected to debate by performing weak measurements in the context of quantum optics experiments \cite{steinberg:PRL:2012,busch:PRL:2013}. So, not only uncertainty relations play a major role in the realm of quantum mechanics, but there is a strong connection to optics since their inception, and are actually implicit in the transition from wave optics to ray optics \cite{harris:JOSA:1969,panuski:PhysTeacher:2016}.

In a simple and elegant manner, uncertainly relations encode valuable information about the maximum accuracy with which two complementary variables can be measured. In other words, the so-called uncertainty principles have proven to be useful signatures to understand the relationship between conjugate Fourier variables, i.e., between the spread of a given variable and its Fourier transform. Besides the aforementioned example, in the particular case of optics we find many processes and phenomena governed by uncertainty relations, which describe conjugate variables (functions or vector fields) that cannot be sharply peaked simultaneously \cite{breitenberger:FoundPhys:1985,stelzer:OptCommun:2000,alonso:JOSAA:2000, santarsiero:JOSAA:2006-1,santarsiero:JOSAA:2006-2, padgett:NJP:2004,padgett:JMO:2008,ricaud:AdvComputMath:2014}. This is the case, for instance, of the resolving power to determine a signal spectrum, which is limited by the duration of the measurement. This is used, in turn, to set the basis for one of the standard measures of laser beam quality \cite{siegman:ProcSPIE:1990,belanger:OptLett:1991,martinezherrero:JOSAA:1991}.

Furthermore, there is also a number of situations that impact the applicability of uncertainty relations. For instance, the standard deviation arising from diffraction through a hard-edged aperture is ill-defined, constraining their application \cite{martinezherrero:OptLett:1993,martinezherrero:OptLett:1995,martinezherrero:JOSAA:2003}. The same is also found when the behavior of the on-axis intensity for highly focused optical fields is analyzed. Yet it would be highly desirable from a practical point of view, for it would allow us to characterize
the beam performance in a rather simple fashion. Recently, the analysis of the properties of optical systems with high numerical aperture has received much attention due to their potential applications in microlithography, image processing, diffractive optics, adaptive optics, or holographic data storage \cite{lindfors:NatPhoton:2007,yzuel:OptLett:2007,zhan:AdvOptPhoton:2009,reid:NatPhoton:2008,sheppard:NatPhoton:2008,dholakia:OptExp:2009,martinezherrero:OptLett:2013,martinezherrero:OptExp:2016, wang:OptCommun:2016,martinezherrero:SciRep:2018,martinezherrero:OSACont:2019,zhang:PRL:2019}. Several performance parameters have thus been considered to compare the focusing behavior of different incident paraxial beams, including the effect of pupil filters \cite{sheppard:JOSAA:1988,sheppard:JMO:1994,martinezcorral:OptLett:2001, martinezcorral:OptExp:2002,sheppard:OptLett:2007,sheppard:OptLett:2008-1,sheppard:OptLett:2008-2,sheppard:OptLett:2011,sheppard:OptExp:2013}.

Here, we tackle the issue and show that the analysis of the global behavior of on-axis power behind a high numerical aperture optical system can be actually specified in terms of an uncertainty principle, which enables a proper measure of localization for the axial power. According to Heisenberg's uncertainty principle, given two conjugate variables, namely $x$ and $u$, their dispersions, measured in terms of the corresponding standard deviations, $\Delta x$ and $\Delta u$, satisfy the inequality
\be
 \left( \Delta x \right)^2 \left( \Delta u \right)^2 \ge 1/4 .
 \label{eq0}
\ee
Only for two Gaussian-distributed variables, this inequality becomes an exact equality.
%
%\be
% \left( \Delta x \right)^2 \left( \Delta u \right)^2 = 1/4 .
% \label{eq00}
%\ee
%
As it is shown below, the behavior of the on-axis intensity after a highly focused optical system can be investigated and understood on an equal footing, where the conjugates variables are the position with respect to the focus of the optical system, $z$, and the polar angle, $\theta$, at the reference Gaussian sphere. Actually, the exact functional form for the incident paraxial beam satisfying the new uncertainty principle here introduced is also specified in terms of a control parameter. According to this control parameter, not only we can design the beam to satisfy the inequality (\ref{eq0}), but it can also be tuned to satisfy the equality condition and even to decrease below $1/4$.

Furthermore, here we also focus on the analysis of the global behavior displayed by the on-axis power at the output of a high numerical aperture optical system. The purpose is to obtain and hence provide a proper location measurement for the axial power content, emphasizing its close and direct relationship with the previously introduced uncertainty principle.

Accordingly, the work is organized as follows. In Sec.~\ref{sec2} we introduce key concepts on the theory of propagation of highly focused beams, and derive the uncertainly principle. In Sec.~\ref{sec3} we investigate the axial region around the focus, where the power content is significant. More specifically, we derive analytical expressions to estimate the axial power content. Finally, the main findings here are summarized in Sec.~\ref{sec4}.

%%%%%%%%%%%%%%%%%%%%%%%%%%%%%%%%%%%%%%%%%%%%%%%%%%%%%%%%%%%%%%%%%%%%%%%%%%%%%%

\section{Uncertainty principle for the axial power content}
\label{sec2}

\begin{figure}[!h]
 \centering
 \includegraphics[width=0.55\columnwidth]{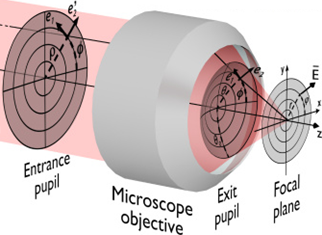}
 \caption{\label{fig1} Sketch of the focusing process through the microscope
  objective showing the reference coordinate systems and variables
  \cite{martinezherrero:OptExp:2014}.}
\end{figure}

Let us consider a monochromatic beam at the entrance pupil of an aplanatic focusing system with a high numerical aperture, as in Fig.~\ref{fig1}. According to the theory on vector field propagation, the electric field at the focal region can be described in terms of the so-called Richards-Wolf integral \cite{richards:PRSLAM:1959},
\be
 {\bf E}(r,\phi,z) = A \int_0^{\theta_0} \!\!\! \int_0^{2\pi} {\bf E}_0(\theta,\varphi) e^{ikr\sin\theta \cos(\phi-\varphi)}
 e^{-ikz\cos\theta} \sin\theta d\theta d\varphi ,
 \label{eq1}
\ee
where $(r,\phi,z)$ are the cylindrical coordinates at the focal area; and $\theta$ and $\varphi$ denote, respectively, the polar and azimuthal angles at the reference Gaussian sphere (see Fig.~\ref{fig1}). The input vector angular spectrum in (\ref{eq1}) is
\be
 {\bf E}_0 = \sqrt{\cos\theta}\ \Big[\! \left({\bf E}_s \cdot {\bf e}_1\right) {\bf e}_1
  + \left({\bf E}_s \cdot {\bf e}'_2\right) {\bf e}_2 \Big] ,
 \label{eq2}
\ee
where ${\bf E}_s$ is the transverse beam distribution at the entrance pupil of the optical system, and the unit vectors
\be
 \begin{array}{rl}
 {\bf e}_1 & = (-\sin\varphi, \cos\varphi, 0) , \\
 {\bf e}_2 & = (\cos\theta \cos\varphi, \cos\theta \sin\varphi, \sin\theta)
 \end{array}
 \label{eq3}
\ee
point along the azimuthal and radial directions, respectively, while ${\bf e}'_2 = (\cos\varphi, \sin\varphi, 0)$ is the projection of ${\bf e}_2$ onto the entrance pupil plane. Regarding the other parameters in (\ref{eq1}), $A$ is a constant, $k$ is the wave number, and $\theta_0 = {\rm max}\{\theta\}$ is the semi-aperture angle, with the numerical aperture being $\mathrm{NA} = \sin\theta_0$.

From the field (\ref{eq1}), the on-axis intensity distribution can be recast as
\be
 I(0,z) = \left\arrowvert {\bf E}(0,z) \right\arrowvert^2 = \left\arrowvert \int_0^{\theta_0} \tilde{\bf F}(\theta) e^{-ikz\cos\theta} \sin\theta d\theta \right\arrowvert^2 ,
 \label{eq4}
\ee
where we define the new vector field
\be
 \tilde{\bf F}(\theta) \equiv A \int_0^{2\pi} {\bf E}_0(\theta,\varphi) d\varphi ,
 \label{eq5}
\ee
depending on the polar angle $\theta$. As it can be noticed, by expressing this field in terms of the new variable $\alpha \equiv \cos \theta$ (with $\alpha_0 = \cos \theta_0$), and the electric field in terms of the dimensionless variable $\bar{z} \equiv kz$, we find that Eq.~(\ref{eq4}) can be written as
\be
 I(0,z) = \left\arrowvert {\bf E}(0,\bar{z}) \right\arrowvert^2
  = \left\arrowvert \int_{\alpha_0}^1 \tilde{\bf F}(\alpha) e^{-i\bar{z}\alpha} d\alpha \right\arrowvert^2 ,
 \label{eq6}
\ee
which stresses the dual relationship between the electric field ${\bf E}(0,\bar{z})$ and the new field $\tilde{\bf F}(\alpha)$, with $\bar{z}$ and $\alpha$ being, in this case, canonically conjugate variables.
The so-called uncertainty principles have proven to be useful signatures to understand the relationship between conjugate Fourier variables, more
specifically, between the dispersion of a given variable and its Fourier
transform.
Here, within the scenario of highly focused fields, we have thus proven that
the same kind of relationship can be found for $\tilde{\bf F}$ whenever this
vector field is continuous and differentiable within the definition domain.

It thus follows, as it is shown with more detail in Appendix~\ref{appA},
that when the continuity and differentiability conditions hold within the
domain $(\alpha_0, 1)$, the field $\tilde{\bf F}$ and its first derivative with respect to the $\alpha$-variable satisfy an uncertainty relation specified by
\be
 \left[ \frac{\displaystyle \int_{\alpha_0}^1 (\alpha - \bar{\alpha})^2 \arrowvert\tilde{\bf F}(\alpha)\arrowvert^2 d\alpha}{I_0} \right]
 \left[ \frac{\displaystyle \int_{\alpha_0}^1 \left\arrowvert \frac{d\tilde{\bf F}(\alpha)}{d\alpha}\right\arrowvert^2 d\alpha}{I_0} \right]
 \ge \frac{1}{4} \left( \frac{B}{I_0} \right)^2 ,
 \label{eq7}
\ee
where
\begin{subequations}
\ba
 I_0 & = & \int_{\alpha_0}^1 \arrowvert\tilde{\bf F}(\alpha)\arrowvert^2 d\alpha ,
 \label{eq8a} \\
 \bar{\alpha} & = & \frac{1}{I_0} \int_{\alpha_0}^1 \alpha \arrowvert\tilde{\bf F}(\alpha)\arrowvert^2 d\alpha ,
 \label{eq8b} \\
 B & = & (1 - \bar{\alpha}) \arrowvert\tilde{\bf F}(1)\arrowvert^2 - (\alpha_0 - \bar{\alpha}) \arrowvert\tilde{\bf F}(\alpha_0)\arrowvert^2 - I_0 .
 \label{eq8c}
\ea
\end{subequations}
It is important to point out that the term on the right of Eq.~(\ref{eq7}) depends on the values of the field at the edges of the entrance pupil [$\tilde{\bf F}(1)$ and $\tilde{\bf F}(\alpha_0)$]. Therefore, the lower bound of this uncertainty relationship depends on the profile and polarization state of the incident beam.
This is an important difference with respect to the usual uncertainty relation, given by Eq~(\ref{eq0}); its physical implications are further
analyzed and discussed below.

To provide further insight on the interpretation of inequality (\ref{eq7}), let us
investigate the physical meaning of the two factors on the left of Eq.~(\ref{eq7}).
In this regard, the first factor clearly represents the standard deviation,
$(\Delta \alpha)^2$, of the vector field $\tilde{\bf F}(\alpha)$.
Concerning the second factor, if the variable $\alpha$ ranged in an infinite interval,
such a factor would describe the second-order moment of the axial intensity. However, in 
general, this moment is not well defined, because the integral that it involves diverges due 
to the finite numerical aperture of the optical system and the discontinuity of
$\tilde{\bf F}$ at the extrema of the definition domain.
Consequently, it is necessary to find a clear physical meaning for this term and its
relationship with the on-axis intensity. This aspect, namely the relationship between the
axial power content and the uncertainty principle here derived, is investigated in more
detail in Sec.~\ref{sec3}.

\begin{figure}[!t]
	\centering
	\includegraphics[width=9cm,height=5cm]{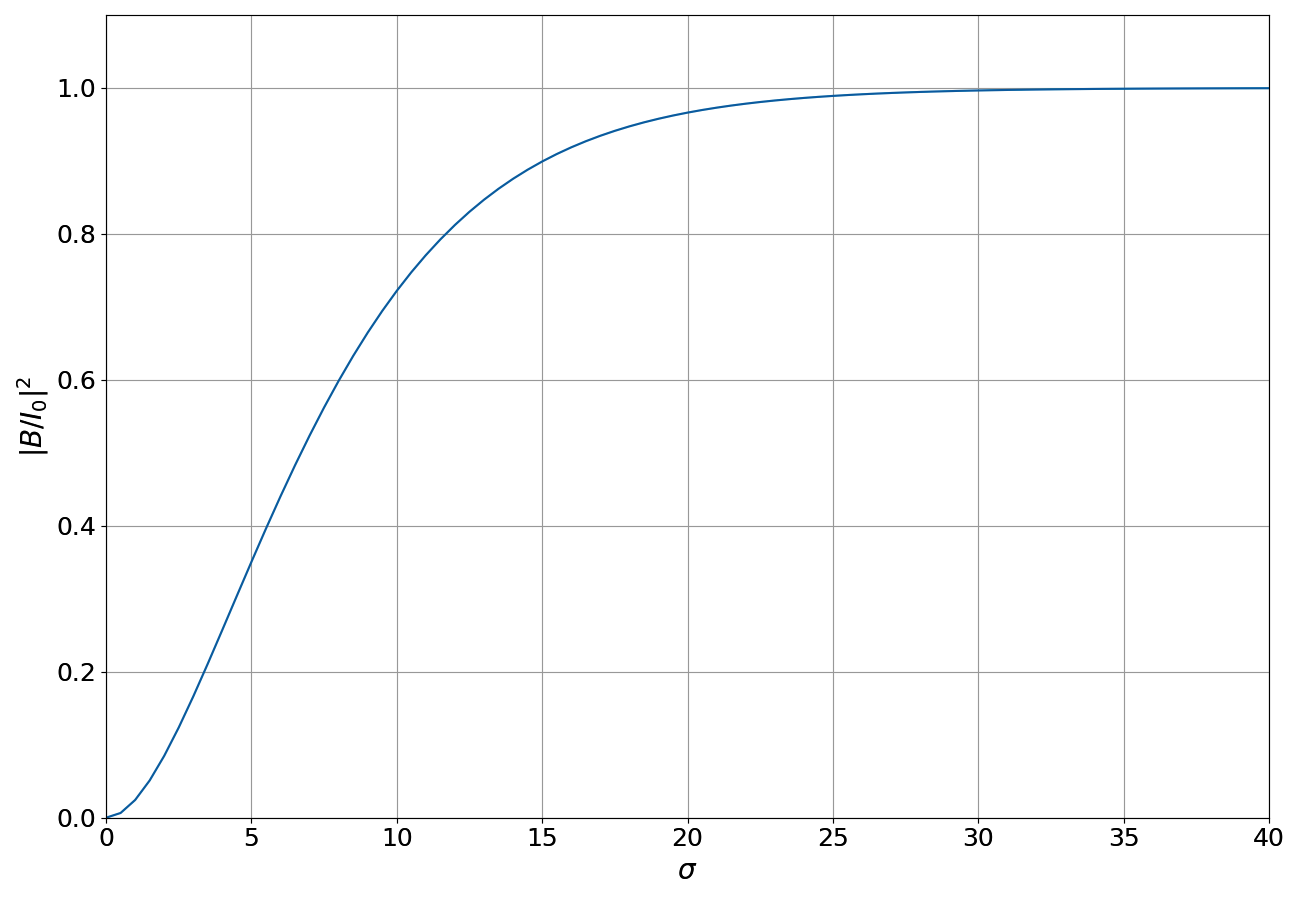}
	\caption{\label{fig2} Dependence of the coefficient $(B/I_0)^2$ [see Eq.~(\ref{eq8c})] on the $\sigma$ parameter for $\mathrm{NA} = 0.9$. As it can be noticed, as $\sigma$ becomes larger, $(B/I_0)^2$ approaches asymptotically the constant value 1 (for practical purposes, this can be assumed to happen beyond $\sigma \ge 25$) and hence it can be considered $\sigma$-independent.}
\end{figure}

Now, let us get back to Eq.~(\ref{eq7}). It can be shown (see Appendix~\ref{appA}) that the equality condition in this relation is reached whenever the incident field is such that the vector field $\tilde{\bf F}(\alpha)$, specified by Eq.~(\ref{eq5}), displays the functional form
\be
 \tilde{\bf F}(\alpha) = e^{- \frac{\sigma}{2} \frac{(\alpha - \bar{\alpha})^2}{(1 - \alpha_0)^2}} {\bf u}
 \label{eq5b}
\ee
in terms of $\alpha$, with ${\bf u}$ being a unitary constant vector and $\sigma$ an arbitrary real, positive parameter.
Note here that, although the modulus of this vector field is a Gaussian
in the $\alpha$-variable, this does not imply that the incident field should
also be Gaussian, which is illustrated below by means of some examples.
Furthermore, it is also worth stressing the fact that from (\ref{eq5b}) it
readily follows
that the term on the right-hand side of the inequality~(\ref{eq7}) is going to be dependent on the parameter $\sigma$. Such a dependence of $(B/I_0)^2$ on $\sigma$ is shown in Fig.~\ref{fig2} for a numerical aperture $\mathrm{NA}=0.9$ ($\alpha_0 = \sqrt{1 - {\rm NA}^2}$). Interestingly, for values $\sigma \leq 25$, the inequality (\ref{eq7}) applied to fields with an associated $\tilde{\bf F}(\alpha)$ field of the form (\ref{eq5b}) becomes
\be
\begin{split}
 \left[ \frac{1}{I_0} \int_{\alpha_0}^1 (\alpha - \bar{\alpha})^2 \arrowvert\tilde{\bf F}(\alpha)\arrowvert^2 d\alpha \right]\!
 \left[ \frac{1}{I_0} \int_{\alpha_0}^1 \left\arrowvert\frac{d\tilde{\bf F}(\alpha)}{d\alpha}\right\arrowvert^2 d\alpha \right]
 \le \frac{1}{4} ,\\
 \label{eq10}
\end{split}
\ee
Moreover, note that for $\sigma \ge 25$, we have $(B/I_0)^2 \sim 1$.

\begin{figure}[!t]
	\centering
	\includegraphics[width=0.5\columnwidth]{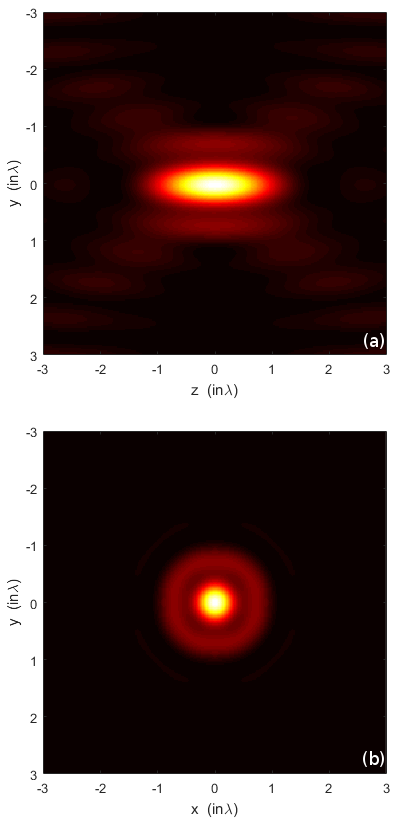}
	\caption{\label{fig3} Irradiance distribution in the focal area (${\rm NA} = 0.9$) for
     an incident azimuthally-polarized paraxial field with topological charge $m=1$ and
     $\sigma=5$: (a) distribution along the $z$-axis for $x=0$, and (b) transverse section
     $z=0$.}
\end{figure}

Given the functional form (\ref{eq5b}), it can readily be seen that different incident fields can be described in terms of it. This is the case, for instance, of an incident paraxial field azimuthally polarized with topological charge $m=1$. To show this, consider the ansatz field
\be
 {\bf E}_s (\theta,\varphi) = g(\theta) e^{i\varphi} {\bf e}_1 ,
 \label{eq9}
\ee
which generates an intensity distribution $I(r,\phi)$ with rotation symmetry around the optical axis, and where the angular distribution function $g(\theta)$ is to be determined. The input vector angular spectrum in this case reads as
\be
 {\bf E}_0 (\theta,\varphi) = \sqrt{\cos \theta} g(\theta) e^{i\varphi} {\bf e}_1 ,
\ee
from which we get
\be
 \tilde{\bf F}(\alpha) = \pi \sqrt{2\alpha} g(\alpha) \bar{\bf u}_c ,
 \label{eq9b}
\ee
with $\bar{\bf u}_c = (-i,1)$. Thus, comparing Eq.~(\ref{eq9b}) with (\ref{eq5b}), we find that the incident field (\ref{eq9}) will be of the kind require to satisfy the above equality condition provided that
\be
 g(\alpha) = \frac{1}{\pi\sqrt{\alpha}}\
  e^{- \frac{\sigma}{2} \frac{(\alpha - \bar{\alpha})^2}{(1 - \alpha_0)^2}} .
\ee
Figure~\ref{fig3} displays the intensity distribution corresponding to this type of incident paraxial beam for ${\rm NA}=0.9$ and $\sigma = 5$. Specifically, panel (a) shows the intensity distribution along the $z$-axis for $x = 0$, while the transverse section at the focus site ($z=0$) is represented in panel (b). In order to get a better and more direct insight from these density plots, $z$ and $x$ are given in units of $\lambda$ instead of considering $k$.

\begin{figure}[!t]
 \centering
 \includegraphics[width=0.5\columnwidth]{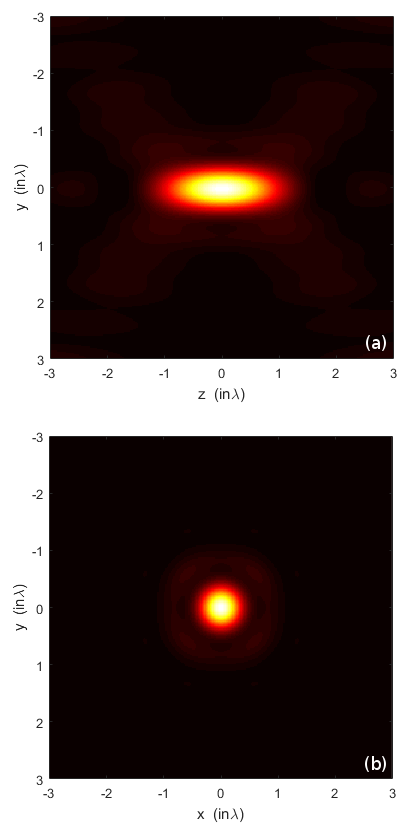}
 \caption{\label{fig4} Irradiance distribution in the focal area (${\rm NA} = 0.9$) for an incident circularly-polarized paraxial field with $\sigma=5$: (a) intensity distribution along the $z$-axis for $x=0$, and (b) transverse section $z=0$.}
\end{figure}

The same procedure can also be followed considering an incident field elliptically polarized instead, i.e.,
\be
{\bf E}_s (\theta,\varphi) = g(\theta) {\bf u}_e ,
\label{eq9c}
\ee
with ${\bf u}_e = (a,bi)$ (with $a^2 + b^2 = 1$), in which case we find
\be
 \tilde{\bf F}(\alpha) = \pi \sqrt{\alpha} (1 + \alpha) g(\alpha) \bar{\bf u}_e ,
\label{eq9d}
\ee
which requires that
\be
g(\alpha) = \frac{1}{\pi\sqrt{\alpha}(1+\alpha)}\
e^{- \frac{\sigma}{2} \frac{(\alpha - \bar{\alpha})^2}{(1 - \alpha_0)^2}} .
\ee
It is important to point out that both types of incident beams lead to the
same function $\tilde{\bf F}(\alpha)$, hence both display the same on-axis intensity distribution.
As before, in Fig.~\ref{fig4} we have plotted the intensity distribution corresponding to an incident paraxial field circularly polarized ($a=b=1/\sqrt{2}$), with ${\rm NA}=0.9$ and $\sigma = 5$. Thus, in panel (a) it is shown the intensity distribution along the $z$-axis for $x = 0$, while the transverse section at the focus site ($z=0$) is given in panel (b). It is worth stressing that these types of beams can be experimentally obtained by using diffractive technology \cite{sheppard:NatPhoton:2008,dholakia:OptExp:2009}.

%%%%%%%%%%%%%%%%%%%%%%%%%%%%%%%%%%%%%%%%%%%%%%%%%%%%%%%%%%%%%%%%%%%%%%%%%%%%%%

\section{Quantification of the axial power content}
\label{sec3}

Besides its inherent fundamental interest, the above uncertainty relation has an also intrinsic applied value in the current scenario of highly focused beams. More specifically, consider we wish to get a fair estimation for the axial power content within a certain region $\Omega$ around the focus, but without involving highly refined and complicated simulations or calculations. To such a purpose, let us define the power-content axial ratio,
\be
 q \equiv \frac{\displaystyle \int_\Omega I(0,\bar{z}) d\bar{z}}{\displaystyle \int_{-\infty}^\infty I(0,\bar{z}) d\bar{z}} ,
 \label{eq11}
\ee
with $\Omega : \bar{z} \in (-kL, kL)$ being the region around the focus ($\bar{z} = 0$), around which the power content takes significant values.
This symmetry condition, though, is not strictly necessary;
in order to ensure maximum on-axis power content, asymmetric boundaries around
the focus, $L_1$ and $L_2$, could also be considered without any loss of
generality, since it is always possible to redefine the origin along the
$z$-axis.
The main purpose now is to find an analytical expression to determine $L$ for a given $q$-value.

As it can be proven (see Appendix~\ref{appB}), $L$ and $q$ are related by means of the inequality
\be
 1 - q \le \Bigg\{ \frac{1}{kL\sqrt{I_0}} \sqrt{\int_{\alpha_0}^1 \left\arrowvert\frac{d\tilde{\bf F}_1(\alpha)}{d\alpha}\right\arrowvert^2 d\alpha}
  + \frac{1}{\sqrt{\pi kLI_0}} \left[ \left\arrowvert\tilde{\bf F}(1) + \tilde{\bf F}(\alpha_0)\right\arrowvert
  + \left\arrowvert\tilde{\bf F}(1) - \tilde{\bf F}(\alpha_0) \right\arrowvert \right] \Bigg\}^2 .
 \label{eq12}
\ee
Here, $\tilde{\bf F}(1)$ and $\tilde{\bf F}(\alpha_0)$ represent the jump discontinuity of the function $\tilde{\bf F}(\alpha)$ at the entrance pupil, while $\tilde{\bf F}_1(\alpha) = \tilde{\bf F}(\alpha) - {\bf G}(\alpha)$, with
\be
 {\bf G}(\alpha) = \frac{(\alpha - \alpha_0) \left[ \tilde{\bf F}(1) - \tilde{\bf F}(\alpha_0) \right]}{1 - \alpha_0} + \tilde{\bf F}(\alpha_0) .
 \label{eq13}
\ee
Accordingly, Eq.~(\ref{eq12}) becomes a profitable analytic tool to determine the size of the region around the focus where a fraction $q$ of the total energy concentrates. To provide further insight on Eq.~(\ref{eq12}), let us define
\be
 kL_0 = \frac{1}{\sqrt{I_0}} \sqrt{\int_{\alpha_0}^1 \left\arrowvert \frac{d\tilde{\bf F}_1(\alpha)}{d\alpha}\right\arrowvert^2 d\alpha}
 + \frac{1}{\sqrt{I_0(1 - \alpha_0)}} \Big[ \left\arrowvert\tilde{\bf F}(1) + \tilde{\bf F}(\alpha_0)\right\arrowvert
  + \left\arrowvert\tilde{\bf F}(1) - \tilde{\bf F}(\alpha_0) \right\arrowvert \Big] .
 \label{eq14}
\ee
As it can be proven (see Appendix~\ref{appB}), within the region $\Omega \in (-aL_0, aL_0)$, the fraction $q$ of axial power content satisfies the inequality $1 - q \le 1/d$, where the right-hand side denotes the maximum bound for the power content, specified in terms of the dimensionless parameter $d$. The relation between this parameter and the bounding factor $a$ is given by
\be
 a = \frac{1}{4\pi} \left[ \sqrt{dQ} (1 - \alpha_0)^{1/4}
 + \sqrt{dQ \sqrt{1 - \alpha_0} + 4\pi\sqrt{d}} \right]^2 ,
 \label{eq15}
\ee
where
\be
 Q = \frac{1}{\sqrt{I_0}} \left[ \left\arrowvert\tilde{\bf F}(1) + \tilde{\bf F}(\alpha_0)\right\arrowvert
  + \left\arrowvert\tilde{\bf F}(1) - \tilde{\bf F}(\alpha_0) \right\arrowvert \right] .
 \label{eq16}
\ee
Notice that, if the vector field $\tilde{\bf F}(\alpha)$ satisfies the constraint
$\tilde{\bf F}(1) = \tilde{\bf F}(\alpha_0) = 0$, then $Q = 0$ and $a = \sqrt{d}$,
thus recovering the usual expression
\be
 kL = \sqrt{d}kL_0 = \sqrt{\frac{d}{I_0}} \sqrt{\int_{\alpha_0}^1 \left\arrowvert\frac{d\tilde{\bf F}(\alpha)}{d\alpha}\right\arrowvert^2 d\alpha} .
 \label{eq17}
\ee
In this case, the value $L$ that guarantees a power axial content of at least 75\% of the total power is given by $L = 2L_0$ (equivalent to considering $d = 4$), i.e., more than 75\% of the axial power content is concentrated in this region. In the general case the region where the axial power is mainly concentrate depends on two terms: the derivative of $\tilde{\bf F}_1$ and the jumps of vector $\tilde{\bf F}$ at the entrance pupil of the objective lens, as stated in Eq.~(\ref{eq14}). The advantage of the preceding definition is that the region were the axial intensity is mainly concentrated can be determined from the knowledge of incident beam (i.e., $\tilde{\bf F}$). In this way, no calculations of the field in focal region are needed. In addition, the ratio $q$ has a clear physical meaning and can be applied to any incident beam. Other advantage of this formalism is that it gives some intuitive guidance for the design of new incident beams in order to obtain a desired axial response \cite{martinezherrero:SciRep:2018,martinezherrero:OSACont:2019}.

Furthermore, taking into account the definition of the vector field ${\bf F}_1(\alpha)$, as specified by Eqs.~(\ref{eq12}) and (\ref{eq13}), the following inequality can be demonstrated (see Appendix~\ref{appC}):
\begin{multline}
 \frac{1}{\sqrt{I_0}} \sqrt{\int_{\alpha_0}^1 \left\arrowvert\frac{d\tilde{\bf F}(\alpha)}{d\alpha}\right\arrowvert^2 d\alpha} \le
 \frac{1}{\sqrt{I_0}} \sqrt{\int_{\alpha_0}^1 \left\arrowvert\frac{d\tilde{\bf F}_1(\alpha)}{d\alpha}\right\arrowvert^2 d\alpha} \\
 + \frac{1}{\sqrt{I_0(1 - \alpha_0)}} \Big[ \left\arrowvert\tilde{\bf F}(1) + \tilde{\bf F}(\alpha_0)\right\arrowvert
 + \left\arrowvert\tilde{\bf F}(1) - \tilde{\bf F}(\alpha_0) \right\arrowvert \Big]
  = kL_0 .
 \label{eq18}
\end{multline}
This inequality allows us to set a connection between the region $\Omega$, where the on axis
energy is mainly concentrated and the uncertainty relationship shown above.
In fact, the uncertainly relationship [Eq.~(\ref{eq7})] can be written in terms of $L_0$, as
\be
 (\Delta \alpha)^2 (kL_0)^2 \ge \frac{1}{4} \left( \frac{B}{I_0} \right)^2 .
 \label{eq19}
\ee

Equation~(\ref{eq19}) provides an interesting physical meaning, the product of standard deviation of variable $\alpha$ and the length $L_0$ where the axial power is mainly concentrated presents a lower limit, that is, both $\Delta \alpha$ and $L_0$ cannot be small simultaneously. This lower limit depends on the jump of the incident field at the entrance pupil of the optical system.
As discussed above, in Sec.~\ref{sec2}, this lower limit might be smaller than $1/4$ for particular choices of the incident paraxial beam, i.e., values of $\sigma$ lower than a threshold value $\sigma_0 \equiv 25$.

\begin{figure}[!h]
 \centering
 \includegraphics[width=0.9\linewidth]{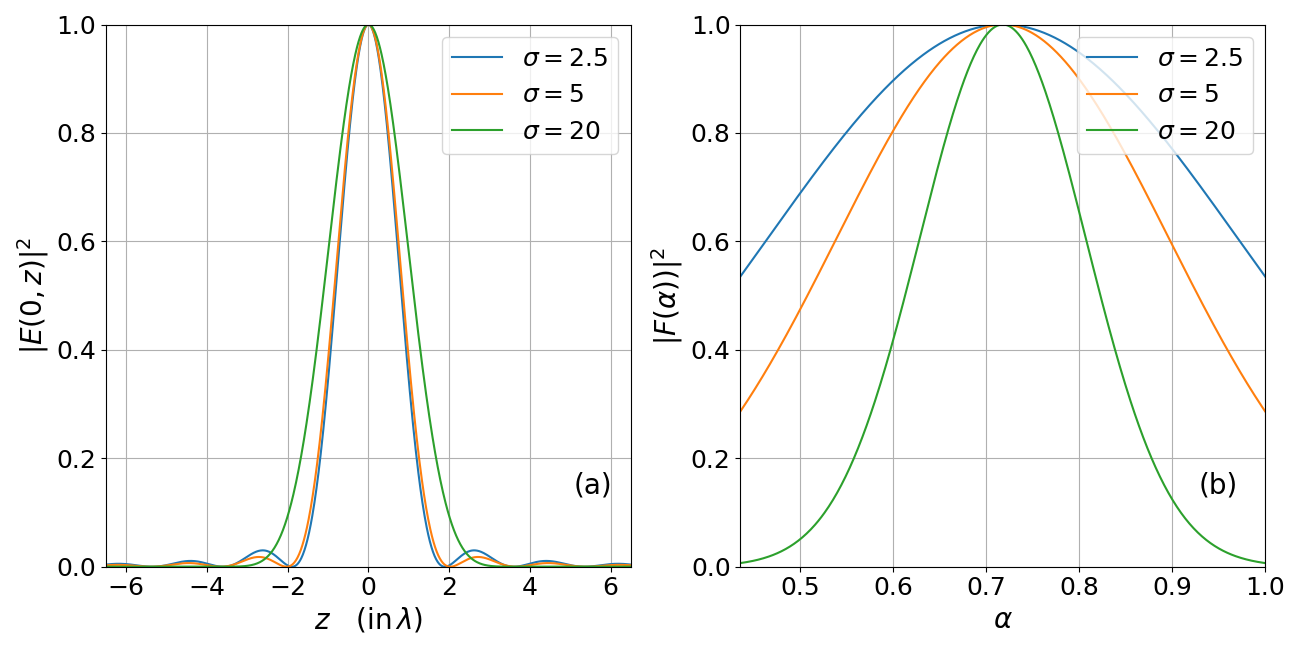}
 \caption{\label{fig5} (a) On-axis intensity distribution $I(0,z)=|{\bf E}(0,z)|^2$ and (b) the corresponding intensity in the dual space $|{\bf F}(\alpha)|^2$ for several values of the $\sigma$ parameter.
 Note that $\bar{z} = kz$. Following Fig.~\ref{fig2}, the values have been chosen in the region where $(B/I_0)^2$  strongly depends on $\sigma$.  As it can be noticed, in this case $(B/I_0)^2$ gets smaller and hence the inequality (\ref{eq7}) is $\sigma$-dependent.}
\end{figure}

\begin{figure}[!h]
 \centering
 \includegraphics[width=0.9\linewidth]{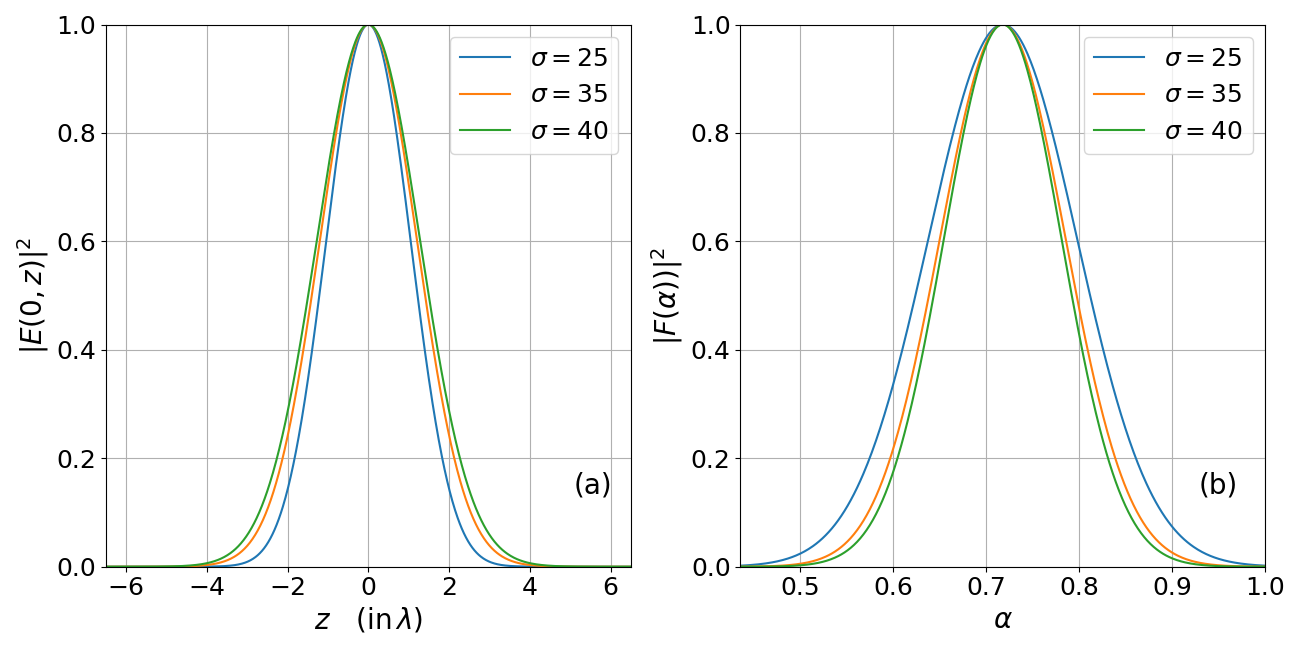}
 \caption{\label{fig6} (a) On-axis intensity distribution $I(0,z)=|{\bf E}(0,z)|^2$ and (b) the corresponding intensity in the dual space $|{\bf F}(\alpha)|^2$ for several values of $\sigma \ge \sigma_0 \equiv 25$, where $(B/I_0)^2$ becomes $\sigma$-independent
  (see Fig.~\ref{fig2}). Note that $\bar{z} = kz$.}
\end{figure}

To illustrate this behavior, we consider the vector $\tilde{\bf F}(\alpha)$ that reaches the equality in the incertitude relation (8).
Figure~\ref{fig5} displays the intensities $|{\bf E}(\bar{z})|^2$ (a) and $|\tilde{\bf F}(\alpha)|^2$ for a series of values $\sigma < \sigma_0$, where there is a strong dependence of $(B/I_0)^2$ on $\sigma$, as seen in Fig.~\ref{fig2}. This implies that the uncertainty principle (\ref{eq19}) is also strongly dependent on such a variable and therefore constitutes a generalization with respect to the usual version of this principle. On the contrary, for $\sigma \ge \sigma_0$, the value of $(B/I_0)^2$ gets very quickly closer and closer to unity and, therefore, Eq.~(\ref{eq7}) becomes $\sigma$-independent, as one would expect for an uncertainty relation. The intensities $|{\bf E}(0,\bar{z})|^2$ and $|{\bf F}(\alpha)|^2$ for different values of $\sigma$ within this regime are both displayed in Fig.~\ref{fig6} [see panels (a) and (b),
respectively].

%%%%%%%%%%%%%%%%%%%%%%%%%%%%%%%%%%%%%%%%%%%%%%%%%%%%%%%%%%%%%%%%%%%%%%%%%%%%%%%

\section{Final remarks}
\label{sec4}

In this work, we introduce a novel analytical procedure to determine the region $\Omega$ in the focal area of a highly focused beam, where the axial power content mainly concentrates. This procedure generalizes the conventional measure of variance that does not apply in this case. Advantageously, this formalism provides us with some intuitive guidance for the design of new pupil functions or incident beams. Furthermore, we demonstrate an uncertainty relation for the size of the area $\Omega$ along the $z$-axis, where the highly-focused intensity and the corresponding incident beam become both more prominent. As it is found, the lower bound of this uncertainty relationship depends on the values of the incident beam at the edge of the entrance pupil of the optical system, which can be smaller than $1/4$. This is one of the main differences with respect to other traditional uncertainty relations. Accordingly, it is possible to simultaneously obtain highly concentrated fields on both the $z$-axis and the entrance pupil. Moreover, we have provided the functional form for the incident beam that reaches the equality in the uncertainty relationship.

To conclude, we would like to note that these results are far more general regarding applicability than the scenario here considered. In our opinion, they might have an important impact in other hot research fields related at present, such as structured light.

%%%%%%%%%%%%%%%%%%%%%%%%%%%%%%%%%%%%%%%%%%%%%%%%%%%%%%%%%%%%%%%%%%%%%%%%%%%

\section*{Funding}

\noindent Spanish Research Agency (AEI) and European Regional Development Fund (ERDF) (FIS2016-75147-C3-1-P, FIS2016-76110-P, PID2019-104268GB-C21, PID2019-104268GB-C22).

%%%%%%%%%%%%%%%%%%%%%%%%%%%%%%%%%%%%%%%%%%%%%%%%%%%%%%%%%%%%%%%%%%%%%%%%%%%

\section*{Disclosures}

\noindent The authors declare no conflicts of interest.

%%%%%%%%%%%%%%%%%%%%%%%%%%%%%%%%%%%%%%%%%%%%%%%%%%%%%%%%%%%%%%%%%%%%%%%%%%%

\appendix

\setcounter{equation}{0}
\def\theequation{A\arabic{equation}}

\section{Derivation of the uncertainty principle}
\label{appA}

Consider a general vector field ${\bf H}(\alpha)$ well behaved in the domain $\alpha \in [\alpha_0,1]$.
The norm of the vector is given by
\be
 \left\| {\bf H}(\alpha) \right\|^2 = \int_{\alpha_0}^1 \left\arrowvert {\bf H}(\alpha) \right\arrowvert^2 d\alpha ,
 \label{eq1-1}
\ee
which is positive definite by definition.
The vector field $\tilde{\bf F}(\alpha)$ here introduced [see Eqs.~(6) and (7)] is of the same form, and hence
its norm is given by (\ref{eq1-1}).
The same holds for any vector field that contains $\tilde{\bf F}(\alpha)$, which allows us to introduce
the inequality \cite{bornwolf-bk}
\be
 \left\| \frac{d\tilde{\bf F}(\alpha)}{d\alpha} - \sigma (\alpha - \bar{\alpha}) \tilde{\bf F}(\alpha) \right\|^2 \ge 0
 \label{eq1-2}
\ee
where
\be
 \bar{a} = \frac{1}{I_0} \int_{\alpha_0}^1 \alpha \arrowvert\tilde{\bf F}(\alpha)\arrowvert^2 d\alpha ,
\ee
with
\be
 I_0 = \int_{\alpha_0}^1 \arrowvert\tilde{\bf F}(\alpha)\arrowvert^2 d\alpha ,
\ee
and $\sigma$ is a real-valued parameter.
The inequality (\ref{eq1-2}) can be recast as
\be
 \sigma^2 \left\| (\alpha - \bar{\alpha}) \tilde{\bf F}(\alpha) \right\|^2 - \sigma \left[ (1 - \bar{\alpha}) \arrowvert\tilde{\bf F}(1)\arrowvert^2 - (\alpha_0 - \bar{\alpha}) \arrowvert\tilde{\bf F}(\alpha_0)\arrowvert^2 - I_0 \right] + \left\| \frac{d\tilde{\bf F}(\alpha)}{d\alpha} \right\|^2
 \ge 0 ,
 \label{eq1-3}
\ee
in terms of the $\sigma$ parameter.
Now, this inequality has to be satisfied by any value of $\sigma$.
Therefore, if we consider the particular case where the equality condition is fulfilled, the real-valuedness
of $\sigma$ leads to the constraint
\be
 B^2 - 4 \left\| \frac{d\tilde{\bf F}(\alpha)}{d\alpha} \right\|^2 \left\| (\alpha - \bar{\alpha}) \tilde{\bf F}(\alpha) \right\|^2 \le 0 ,
 \label{eq1-4}
\ee
with
\be
 B = (1 - \bar{\alpha}) \arrowvert\tilde{\bf F}(1)\arrowvert^2 - (\alpha_0 - \bar{\alpha}) \arrowvert\tilde{\bf F}(\alpha_0)\arrowvert^2 - I_0 .
\ee
As it is easily seen, by rearranging terms, the constraint (\ref{eq1-4}) can be rewritten in the
form of an uncertainty relation.

It is worth noting that the equality condition in (\ref{eq1-3}) is satisfied for the particular
case that the vector field $\tilde{\bf F}(\alpha)$ displays the functional form
\be
 \tilde{\bf F}(\alpha) = A \exp \left[ -\frac{\sigma (\alpha - \bar{\alpha})^2}{2(1 - \alpha_0)^2} \right] {\bf u} ,
 \label{eq1-5}
\ee
with ${\bf u}$ being a constant, unitary vector, $\sigma$ a real, positive parameter, and $\bar{\alpha} = (1 + \alpha_0)/2$.

%%%%%%%%%%%%%%%%%%%%%%%%%%%%%%%%%%%%%%%%%%%%%%%%%%%%%%%%%%%%%%%%%%%%%%%%%%%

\setcounter{equation}{0}
\def\theequation{B\arabic{equation}}

\section{On-axis power content}
\label{appB}

Let us consider the auxiliary vectors $\tilde{\bf F}_1(\alpha)$ and ${\bf G}(\alpha)$, with
\begin{subequations}
\ba
 \tilde{\bf F}_1(\alpha) & = & \tilde{\bf F}(\alpha) - {\bf G}(\alpha) , \\
 {\bf G}(\alpha) & = & \frac{(\alpha - \alpha_0) \left[ \tilde{\bf F}(1) - \tilde{\bf F}(\alpha_0) \right]}{1 - \alpha_0} + \tilde{\bf F}(\alpha_0) .
\ea
\end{subequations}
In terms of these vector, the on-axis field can be recast as
\be
 {\bf E}(0,\bar{z}) = {\bf E}_1(0,\bar{z}) + {\bf E}_2(0,\bar{z}) ,
 \label{eq2-1}
\ee
with
\begin{subequations}
\ba
 {\bf E}_1(0,\bar{z}) & = & \int_{\alpha_0}^1 \tilde{\bf F}_1(\alpha) e^{-i\bar{z}\alpha} d\alpha ,
 \label{eq2-2} \\
 {\bf E}_2(0,\bar{z}) & = & \int_{\alpha_0}^1 {\bf G}(\alpha) e^{-i\bar{z}\alpha} d\alpha ,
 \label{eq2-3}
\ea
\end{subequations}
and $\bar{z} = kz$.
We can then establish the following inequality
\be
 \int_{\mathbb{R}-\Omega} |{\bf E}(0,\bar{z})|^2 d\bar{z} \le \left[ \sqrt{\int_{\mathbb{R}-\Omega} |{\bf E}_1(0,\bar{z})|^2 d\bar{z}}
 + \sqrt{\int_{\mathbb{R}-\Omega} |{\bf E}_2(0,\bar{z})|^2 d\bar{z}} \right]^2
 \label{eq2-4}
\ee
for the on-axis power content, with $\Omega : \bar{z} \in (-kL,kL)$.

The second-order intensity moment for ${\bf E}_1(0,z)$ is mathematically well behaved
[note that $\tilde{\bf F}_1(1) = \tilde{\bf F}_1(\alpha) = 0$] and a bound for it can be
found in terms of the dimensionless variable $\bar{z}$, which reads as
\be
 \int_{\mathbb{R}-\Omega} |{\bf E}_1(0,\bar{z})|^2 d\bar{z} \le
  \frac{\displaystyle 2\pi \int_{\alpha_0}^1 \left\arrowvert \frac{d\tilde{\bf F}_1(\alpha)}{d\alpha} \right\arrowvert^2 d\alpha}{(k L)^2} .
 \label{eq2-6}
\ee
Proceeding now with ${\bf E}_2(0,z)$, as given by Eq.~(\ref{eq2-3}), this field is first recast as
\be
 {\bf E}_2(0,\bar{z}) = \frac{(1 - \alpha_0)}{2} \left\{ \left[ \tilde{\bf F}(1) + \tilde{\bf F}(\alpha_0) \right]
  {\rm sinc}\left[ \frac{\bar{z}(1 - \alpha_0)}{2} \right]
  - i \left[ \tilde{\bf F}(1) - \tilde{\bf F}(\alpha_0) \right] j_1 \left[ \frac{\bar{z}(1 - \alpha_0)}{2} \right] \right\}^2 ,
  \label{eq2-7}
\ee
where $j_1$ is the spherical Bessel function of the first kind.
Taking into account the asymptotic behavior displayed by the functions $j_1$ and ${\rm sinc}$,
we can also readily find a bound for the on-axis power content associated with ${\bf E}_2(0,\bar{z})$,
which is given by
\ba
 \int_{\mathbb{R}-\Omega} |{\bf E}_2(0,\bar{z})|^2 d\bar{z} \le \left\{ \frac{\displaystyle \sqrt{2} \left[ \left\arrowvert \tilde{\bf F}(1) - \tilde{\bf F}(\alpha_0) \right\arrowvert
  + \left\arrowvert \tilde{\bf F}(1) - \tilde{\bf F}(\alpha_0) \right\arrowvert \right]}{\displaystyle \sqrt{kL}} \right\}^2 . \nonumber \\
 \label{eq2-8}
\ea

If Eqs.~(\ref{eq2-6}) and (\ref{eq2-8}) are now substituted into Eq.~(\ref{eq2-4}), we obtain
\be
 \int_{\mathbb{R}-\Omega} \!\!\! |{\bf E}(0,\bar{z})|^2 d\bar{z} \le
   \left\{ \frac{\sqrt{2} \left[ \left\arrowvert \tilde{\bf F}(1) - \tilde{\bf F}(\alpha_0) \right\arrowvert
  + \left\arrowvert \tilde{\bf F}(1) - \tilde{\bf F}(\alpha_0) \right\arrowvert \right]}{\sqrt{kL}}
  + \frac{\sqrt{2\pi \int_{\alpha_0}^1 \left\arrowvert \frac{d\tilde{\bf F}_1(\alpha)}{d\alpha} \right\arrowvert^2 d\alpha}}{kL} \right\}^2 .
 \label{eq2-9}
\ee
Inserting Eq.~(\ref{eq2-9}) into Eq.~(11), we finally get
\be
 1 - q \le \left\{ \frac{\sqrt{\int_{\alpha_0}^1 \left\arrowvert \frac{d\tilde{\bf F}_1(\alpha)}{d\alpha} \right\arrowvert^2 d\alpha}}{kL\sqrt{I_0}}
  + \frac{\left[ \left\arrowvert \tilde{\bf F}(1) + \tilde{\bf F}(\alpha_0) \right\arrowvert
  + \left\arrowvert \tilde{\bf F}(1) - \tilde{\bf F}(\alpha_0) \right\arrowvert \right]}{\sqrt{\pi kL I_0}} \right\}^2 .
 \label{eq2-10}
\ee

In order to simplify the notation, first we introduce the new variables
\begin{subequations}
\ba
 P & = & \sqrt{\frac{\displaystyle \int_{\alpha_0}^1 \left\arrowvert \frac{d\tilde{\bf F}_1(\alpha)}{d\alpha} \right\arrowvert^2 d\alpha}{I_0}} ,
 \label{eq2-11a} \\
 Q & = & \frac{1}{\sqrt{I_0}} \left[ \left\arrowvert \tilde{\bf F}(1) - \tilde{\bf F}(\alpha_0) \right\arrowvert
  + \left\arrowvert \tilde{\bf F}(1) + \tilde{\bf F}(\alpha_0) \right\arrowvert \right] .
 \label{eq2-11b}
\ea
\end{subequations}
Accordingly, Eq.~(\ref{eq2-10}) can be recast as
\be
 1 - q \le \left(\frac{P}{kL} + \frac{Q}{\sqrt{\pi kL}}\right)^2 .
 \label{eq2-12}
\ee
If we now define
\be
 L = aL_0 ,
 \label{eq2-13}
\ee
with
\be
 kL_0 = P + \frac{Q}{\sqrt{1 - \alpha_0}} ,
 \label{eq2-14}
\ee
the following bound holds,
\be
 \frac{P}{kL} + \frac{Q}{\sqrt{\pi kL}} \le \frac{1}{a} + \frac{\sqrt{Q} (1 - \alpha_0)^{1/4}}{\sqrt{\pi a}} .
 \label{eq2-15}
\ee
The right-hand side of Eq.~(\ref{eq2-15}) can be recast in terms of a dimensionless bounding parameter, $d$,
which serves to the purpose of limiting the power content (left-hand side).
Accordingly, if we introduce the following definition
\be
 \frac{1}{a} + \frac{\sqrt{Q} (1 - \alpha_0)^{1/4}}{\sqrt{\pi a}} = \frac{1}{\sqrt{d}} ,
 \label{eq2-17}
\ee
the value we obtain for $a$ is
\be
 a = \frac{1}{4\pi} \left[ \sqrt{dQ} (1 - \alpha_0)^{1/4} + \sqrt{dQ \sqrt{1 - \alpha_0} + 4\pi\sqrt{d}} \right]^2 .
 \label{eq2-16}
\ee
in terms of the new dimensionless variable $d$.
Thus, substituting Eq.~(\ref{eq2-17}) into (\ref{eq2-15}), and the resulting expression into
Eq.~(\ref{eq2-12}), we finally obtain the inequality
\be
 1 - q \le \frac{1}{d} ,
\ee
which is a more convenient, simplified version of the inequality (\ref{eq2-10}).

%%%%%%%%%%%%%%%%%%%%%%%%%%%%%%%%%%%%%%%%%%%%%%%%%%%%%%%%%%%%%%%%%%%%%%%%%%%

\setcounter{equation}{0}
\def\theequation{C\arabic{equation}}

\section{Proof of the inequality (\ref{eq18})}
\label{appC}

We rewrite Eq.~(13) as
\be
 \tilde{\bf F}(\alpha) = \tilde{\bf F}_1(\alpha) + {\bf G}(\alpha) ,
 \label{eq3-1}
\ee
then
\ba
 \sqrt{\int_{\alpha_0}^1 \left\arrowvert \frac{d\tilde{\bf F}(\alpha)}{d\alpha} \right\arrowvert^2 d\alpha} & \le &
 \sqrt{\int_{\alpha_0}^1 \left\arrowvert \frac{d\tilde{\bf F}_1(\alpha)}{d\alpha} \right\arrowvert^2 d\alpha}
 + \sqrt{\int_{\alpha_0}^1 \left\arrowvert \frac{d{\bf G}(\alpha)}{d\alpha} \right\arrowvert^2 d\alpha}
 \nonumber \\
 & = &
 \sqrt{\int_{\alpha_0}^1 \left\arrowvert \frac{d\tilde{\bf F}_1(\alpha)}{d\alpha} \right\arrowvert^2 d\alpha}
 + \frac{\left\arrowvert \tilde{\bf F}(1) - \tilde{\bf F}(\alpha_0) \right\arrowvert}{\sqrt{1 - \alpha_0}}
 \nonumber \\
 & \le &
 \sqrt{\int_{\alpha_0}^1 \left\arrowvert \frac{d\tilde{\bf F}_1(\alpha)}{d\alpha} \right\arrowvert^2 d\alpha}
 + \frac{\left\arrowvert \tilde{\bf F}(1) - \tilde{\bf F}(\alpha_0) \right\arrowvert
   + \left\arrowvert \tilde{\bf F}(1) + \tilde{\bf F}(\alpha_0) \right\arrowvert}{\sqrt{1 - \alpha_0}}
 \nonumber \\
 & \le & kL_0 .
 \label{eq3-2}
\ea

%%%%%%%%%%%%%%%%%%%%%%%%%%%%%%%%%%%%%%%%%%%%%%%%%%%%%%%%%%%%%%%%%%%%%%%%%%%
%%%%%%%%%%%%%%%%%%%%%%%%%%%%%%%%%%%%%%%%%%%%%%%%%%%%%%%%%%%%%%%%%%%%%%%%%%%

%\bibliography{references}

\end{document}